%% file: ms.tex
\input defs.tex

\documentclass[12pt,preprint]{aastex}
\usepackage{emulateapj5,natbib}
\begin{document}

\title{Effect of Internal Flows on Sunyaev-Zeldovich Measurements of
  Cluster Peculiar Velocities} 
\author{Daisuke Nagai\altaffilmark{1,2}, 
        Andrey V. Kravtsov\altaffilmark{1,2}, 
        Arthur Kosowsky\altaffilmark{1,3} 
} 

\altaffiltext{1}{Center for Cosmological Physics, University of Chicago,
Chicago IL 60637}

\altaffiltext{2}{Department of Astronomy and Astrophysics,
University of Chicago, 5640 S. Ellis Ave, Chicago IL 60637}

\altaffiltext{3}{Department of Physics and Astronomy,
  Rutgers University, 136 Frelinghuysen Road, Piscataway, NJ
  08854-8019}

\begin{abstract}
  Galaxy clusters are potentially powerful probes of the large-scale
  velocity field in the Universe because their peculiar velocity can
  be estimated directly via the kinematic Sunyaev-Zeldovich effect
  (kSZ).  Using high-resolution cosmological simulations of an
  evolving cluster of galaxies, we evaluate how well the average
  velocity obtained via a kSZ measurement reflects the actual cluster
  peculiar velocity. We find that the internal velocities in the
  intracluster gas are comparable to the overall cluster peculiar
  velocity, 20 to 30\% of the sound speed even when a cluster is
  relatively relaxed. Nevertheless, the velocity averaged over the kSZ
  map inside a circular aperture matched to the cluster virial region
  provides an unbiased estimate of a cluster's radial peculiar
  velocity with a dispersion of 50 to 100 km/s, depending on the line
  of sight and dynamical state of the cluster.  This dispersion puts a
  lower limit on the accuracy with which cluster peculiar velocity can
  be measured. Although the dispersion of the average is relatively
  small, the velocity distribution is broad; regions of low signal
  must be treated with care to avoid bias.  We discuss the extent to
  which systematic errors might be modelled, and the resulting
  limitations on using galaxy clusters as cosmological velocity
  tracers.
\end{abstract}



\section{Introduction}
\label{sec:intro}

Accurate maps of the cosmic velocity field would provide powerful
constraints on the formation of structure in the Universe. A velocity
field can be compared directly with measured galaxy density field,
testing whether the fundamental picture of gravitational collapse is
correct \citep{branchini01,dacosta98,dore02}, or with velocity fields
predicted by cosmological models, providing useful constraints on
cosmological parameters and an independent measurement of the bias
parameter on a range of scales. Indeed, intensive theoretical effort
in the last decade produced very accurate predictions of various
properties of the velocity fields from the linear to the highly
nonlinear regime
\citep[e.g.,][]{jing98,freudling99,juszkiewicz99,colin00}.  However,
despite clear theoretical importance, velocity surveys have been
overshadowed by recent redshift surveys because peculiar velocities
are far more difficult to measure than redshifts.  Existing velocity
surveys are thus more prone to systematic errors than redshift
surveys.  The main difficulty in measuring peculiar velocity is, of
course, obtaining the distance to an object, which is necessary if the
peculiar velocity is obtained by subtracting the Hubble flow velocity
from a measured redshift velocity.  Since distance errors tend to
increase with distance, the reliability of redshift-based peculiar
velocity surveys inevitably degrades with distance, making
cosmological tests on large scales difficult.

Several groups have proposed using clusters of galaxies as tracers of
the cosmic velocity field
\citep{haehnelt96,lange98,kashlinsky00,aghanim01,peel02}. The peculiar
velocity of a cluster can be measured in a single step, independent of
the cluster distance, via the kinematic Sunyaev-Zeldovich effect
\citep[SZE :][]{zeldovich69,sunyaev80}; for recent reviews see
\citet{birkinshaw99} and \citet{carlstrom02}.  The electrons of the hot ionized
gas in clusters Compton-scatter passing microwave background photons,
resulting in a frequency-dependent redistribution of the photon
energies (the thermal SZ effect).  A smaller but measurable blackbody
distortion arises from the Doppler shift due to bulk motion of the
electrons with respect to the rest frame of the CMB photons (the
kinematic SZ effect, hereafter kSZ). In principle, these effects can
be distinguished by their spectral signatures. With sufficiently
high angular resolution observations the kSZ signal can be
distinguished from intrinsic microwave background temperature
fluctuations because galaxy clusters subtend angular scales small
compared to characteristic primordial fluctuations. First measurements
of the kSZ signal in clusters have been reported \citep{holzapfel97},
and several studies of the systematic errors related to disentangling
the kSZ signal from other microwave fluctuations have been performed
\citep{haehnelt96,aghanim01,diego02}.

If clusters were simple spherical objects with negligible internal
structure and motions (as has been assumed in several previous papers
investigating the effect), then measurements of the kinematic SZ
effect would provide a relatively straightforward method for measuring
the cluster peculiar velocity. However, we know from high-resolution
X-ray observations
\citep[i.e.,][]{markevitch00,vikhlinin01,sun02,mazzotta02} and
numerical simulations \citep[i.e.,][]{norman99,ricker01} that galaxy
clusters are actually complex objects with significant internal flows
driven by frequent mergers. The observed kinematic SZ signal, which
measures the density-weighted peculiar velocity of gas, will thus be
some average over the bulk velocity and the internal velocities, which
raises the question of how accurately an observed SZ signal will
reflect the actual peculiar velocity of the cluster. Potential
systematic errors can arise because the gas itself does not
necessarily reflect the bulk velocity of the matter.

In this paper we use high-resolution simulations of a galaxy cluster
formed in a $\Lambda$CDM cosmological model to investigate the
systematic errors that can arise from the complex internal structure
and motion in estimates of the cluster velocity via the kSZ effect.
In $\S$2, we describe the cluster simulation used in our analysis.  In
$\S$3, we illustrate the structure and magnitude of internal flows
within the cluster.  $\S$4 explains how to estimate the peculiar
velocity from kSZ maps, discusses definitions of peculiar velocity and
analyzes the systematic errors in the kSZ estimate. We summarize our
results and discuss related issues in the final section.

\section{Numerical Simulations}
\label{sec:numdetail}

We analyze a high-resolution cluster simulation performed using the
Adaptive Refinement Tree (ART) $N$-body$+$gasdynamics code
\citep{kravtsov99, kravtsov02}.  ART is an Eulerian code designed to
achieve high spatial resolution by adaptively refining regions of
interest, such as high-density regions or regions of steep gradients
in gas properties, and has good shock-capturing characteristics.  The
code can capture discontinuities in gas properties within $\sim 1-2$
grid cells without using artificial viscosity, which makes it
well-suited for studying sharp features such as shock fronts and
studying velocity fields.  At the same time, adaptive refinement in
space and (non-adaptive) refinement in mass \citep{klypin01} allows us
to reach the large dynamic range required for high-resolution
self-consistent simulations of cluster evolution in a cosmological
setting.

The effects of magnetic fields, gas cooling, stellar feedback, and
thermal conduction are not included in these simulations. As we show
below, the systematic errors in the measurements of the radial
peculiar velocity of clusters from kSZ observations arise
mainly from the proximity of merging sub-clumps and the internal
turbulent motion of cluster gas induced by mergers.  These dynamical
processes are expected to be largely unaffected by the neglected
physical processes. More importantly, properly assessing the effect of
gas flows requires simulations of cluster formation in a realistic
cosmological setting, taking into account non-spherical accretion of
matter along filaments along with major and minor mergers which induce
the internal gas motions. The simulation analyzed here follows
formation of a galaxy cluster from CDM initial conditions and has both
very high mass and spatial dynamic range to model accurately the
small-scale dynamics of dark matter and gas.

Specifically, we analyze a simulation of a cluster of intermediate
mass in a $\Lambda$CDM model with $\Omega_m=1-\Omega_{\Lambda}=0.3$,
$\Omega_b=0.043$, $h=0.7$ and $\sigma_8=0.9$, where the Hubble
constant is defined as $100h{\ \rm km\ s^{-1}\ Mpc^{-1}}$, and
$\sigma_8$ is the rms density fluctuation on $8h^{-1}$~Mpc scales. The
simulation used a base 128$^3$ uniform grid and 7 levels of mesh
refinement in a computational box of $80h^{-1}$~Mpc with a peak formal
resolution of $5\hkpc$ and mass resolution (particle mass) of
$2.7\times 10^8h^{-1}{\rm M_{\odot}}$.  At $z=0$, the cluster has
virial mass of $M_{340}=2.4\times 10^{14}\hmsol$ and virial radius of
$R_{340}=1.26$\hmpc, where the subscripts indicate the
cosmology-dependent virial overdensity with respect to the mean
density of the Universe.  This $\Lambda$CDM cluster simulation was
recently used in the analysis of ``cold fronts'' in clusters
\citep{nagai02a}, and further numerical details of the cluster
simulation can be found there.

\section{Internal flows}
\label{sec:flows}

The main question we want to answer is the extent to which the
internal flows intrinsic to clusters will degrade measurements of
cluster peculiar velocities inferred from the kSZ effect.  We start by
illustrating the structure and magnitude of internal flows using the
high-resolution simulation of a galaxy cluster described above.

\myputfigure{Plot_LowRes/f1.ps}{3.0}{0.43}{-5}{+5} 
\figcaption{ \footnotesize
  A gas velocity map overlayed on the emission-weighted temperature
  map of the cluster at $z=0$. The gas velocity is the average
  projected velocity in a $78h^{-1}$Mpc slice centered on the central
  density peak.  The comoving size of the region is 0.82$h^{-1}$Mpc.
  The length of the thick vertical vector in the bottom-left corner
  corresponds to 300 \kms; the temperature maps are color-coded on a
  $\log_{10}$ scale in units of degrees Kelvin.
\label{fig1}
\vskip10pt
}

Figure~\ref{fig1} shows velocity maps of gas overlayed on the
emission-weighted temperature map of a relatively relaxed cluster at
$z=0$.  The velocity map shows the average gas velocity in a 78$\hkpc$
slice centered on the central density peak; the comoving size of the
displayed region is 0.82$h^{-1}$ Mpc.  The velocity map shows internal
bulk motions of gas at a level of $\sim 200$\kms within this relaxed
cluster. The typical flow velocities are $\sim 10-30\%$ of the sound speed 
in the cluster core.  Gas motions of a similar magnitude
in the cores of relaxed clusters are implied by the Chandra
observations \citep{markevitch02}.  Note the merging sub-clump with
velocity $\gtrsim 1000$\kms on the left side of the main cluster.
Regions of fast-moving gas are generally present even in ``relaxed''
clusters due to ongoing minor mergers.  These prominent internal
motions are potential sources of bias in the peculiar velocity
estimate.

\myputfigure{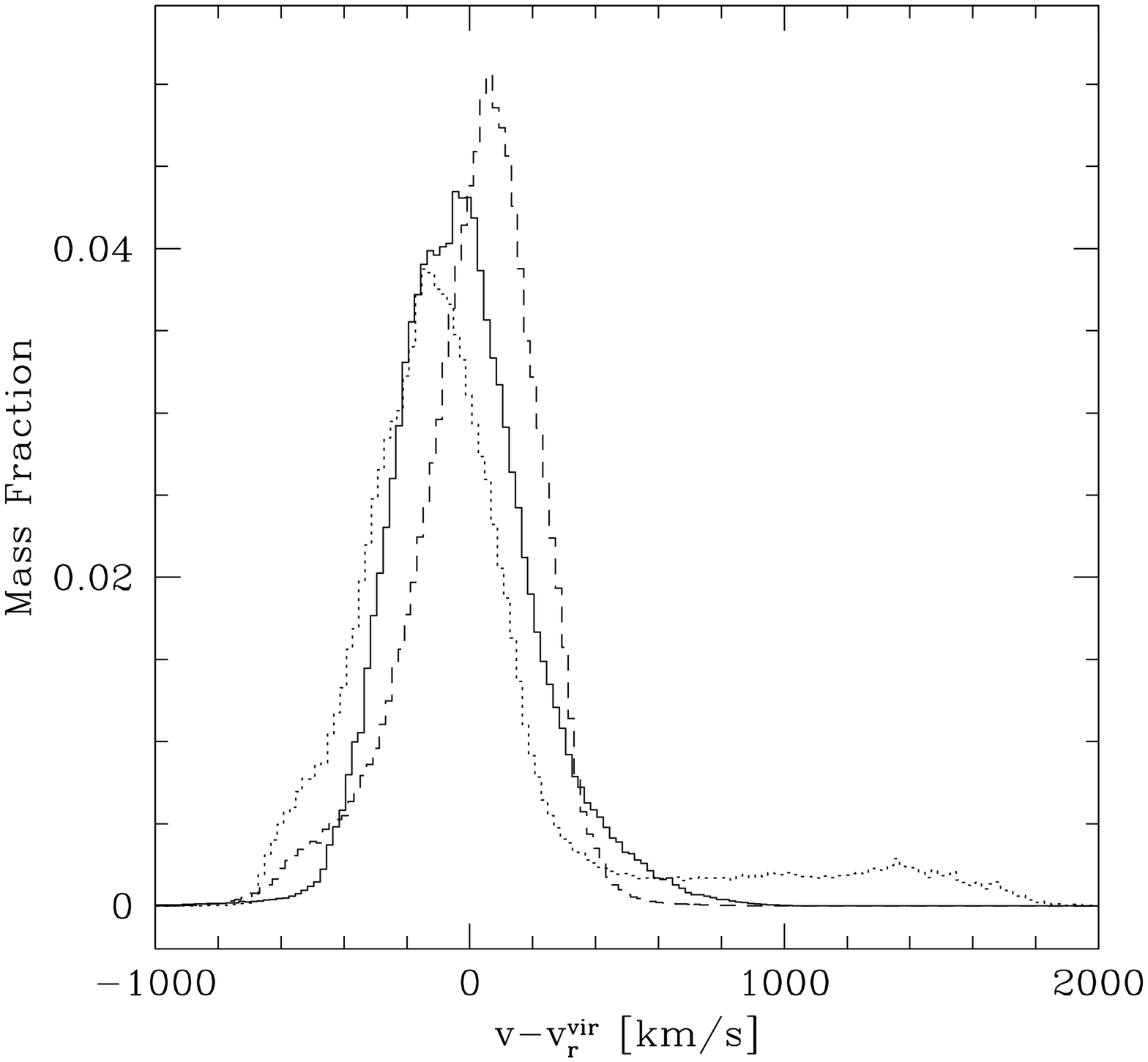}{3.2}{0.50}{-15}{-0}
\figcaption{ \footnotesize 
  The distribution of the gas velocity component along three
  orthogonal projections within the virial radius of the cluster at
  $z=0$.  (Solid, dotted, and dashed lines correspond to the left,
  middle, and right panel projections in Fig.~\ref{fig3} below.)
\label{fig2}
\vskip10pt
}

Figure~\ref{fig2} shows the distribution of the gas velocity component
along three orthogonal projections within the virial radius of the
cluster at $z=0$.  The velocity distributions are Gaussian with
velocity dispersion of $\sim$200\kms, except for one case where a
merging sub-clump has a line-of-sight velocity $\gtrsim 1000$\kms (see
also Fig.~\ref{fig1}).  The symmetric nature of the velocity
distribution indicates that the kSZ velocity estimate will generally
reflect the true peculiar velocity, but that merging high-velocity
substructures along the line of sight can bias the measurements, even
given an exact recovery of the kSZ signal.  Our simulation suggests
that such substructure occurs in a non-negligible fraction of
clusters.

\section{Peculiar Velocities From the Kinematic SZ Effect}
\label{sec:ksz}

Can we measure the radial peculiar velocity of clusters using the kSZ
effect accurately enough to learn about cosmology?  In the previous
section, we have demonstrated that internal gas motion and fast-moving
merging sub-clumps are generally present even in the relaxed cluster
with velocities comparable to the overall cluster peculiar velocity.
So, sizable systematic errors are expected in the peculiar velocity
estimate via kSZ effect, even if a recovery of the kSZ signal is
perfect. This raises a question of how accurately the kSZ signal will
reflect the actual peculiar velocity of cluster.  To address this
issue, we assume the exact recovery of the kSZ signal and construct
noiseless kSZ distortion maps from the high-resolution cluster
simulation at different epochs and along three orthogonal directions.
We then estimate radial peculiar velocities from these maps and
compare the value to the radial peculiar velocity of a cluster defined
by the smoothed dark matter distribution.

Note that the question addressed here is separate from the question
of how well the kSZ signal can actually be extracted from a realistic
map, both technically and conceptually.  Other uncertainties,
particularly confusion with the lensing signal and point source
contamination, will give systematic errors in the estimate of the kSZ
signal itself, which will propagate into peculiar velocity
estimates. These error sources will be considered elsewhere. The error
we estimate below is the {\em irreducible} peculiar velocity error due to 
internal motions of the intracluster gas. 

\subsection{Simulated kSZ Maps}
\label{sec:map}

The kSZ effect arises due to Compton scattering of CMB photons by
ionized gas in motion with respect to the rest frame of the CMB. The
resulting induced temperature fluctuation is
\beq 
-{\Delta T\over T}(\vec{x}) = b(\vec{x}) \equiv \sigma_T \int dl \: n_e v_r,
\label{eq:b_def}
\eeq
where $\sigma_T$ is the Thompson cross section, $n_e$ is the electron
number density, $v_r$ is the velocity component along the line of
sight (in units of the speed of light), the integration is along the
line of sight, and $\vec{x}$ represents directions on the (flat) sky.
We generate kSZ maps of the simulated cluster from uniform density and
velocity grids centered on the minimum of cluster potential by
numerically integrating Eq.~\ref{eq:b_def}. The density and velocity
grids, in turn, were constructed using interpolation from the original
mesh refinement structure. We then model measurements at a given
resolution by convolving the maps with a Gaussian beam profile
$A_\sigma(r)$ of appropriate width.  Throughout this paper, we will
assume a FWHM beamwidth of 1 arcminute.  While individual clusters
will be observed with higher resolution than this, it is unlikely that
large SZ cluster surveys useful for cosmology will obtain
significantly higher resolution in the near future. Higher resolution
would not significantly alter our conclusions since the dominant
systematic effects are unrelated to whether fine-scale structure is
resolved in the maps.

\begin{figure*}[t]
\centerline{ 
\epsfysize=3.6truein  \epsffile{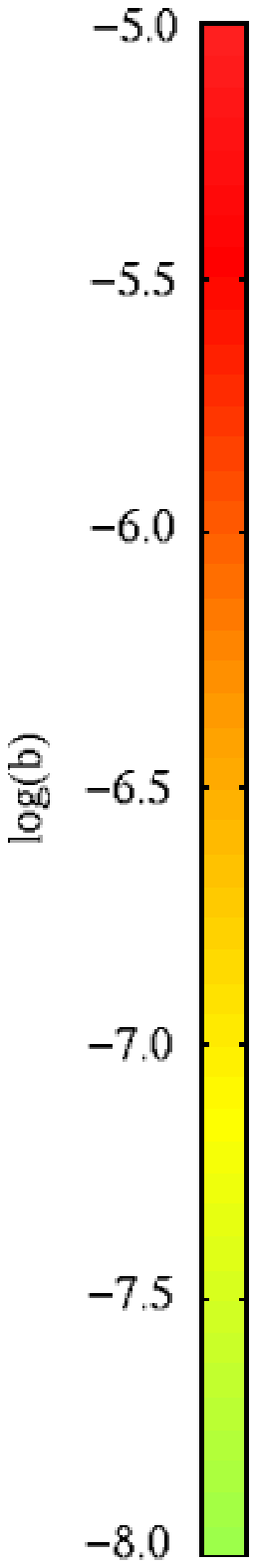} 
\epsfysize=3.6truein  \epsffile{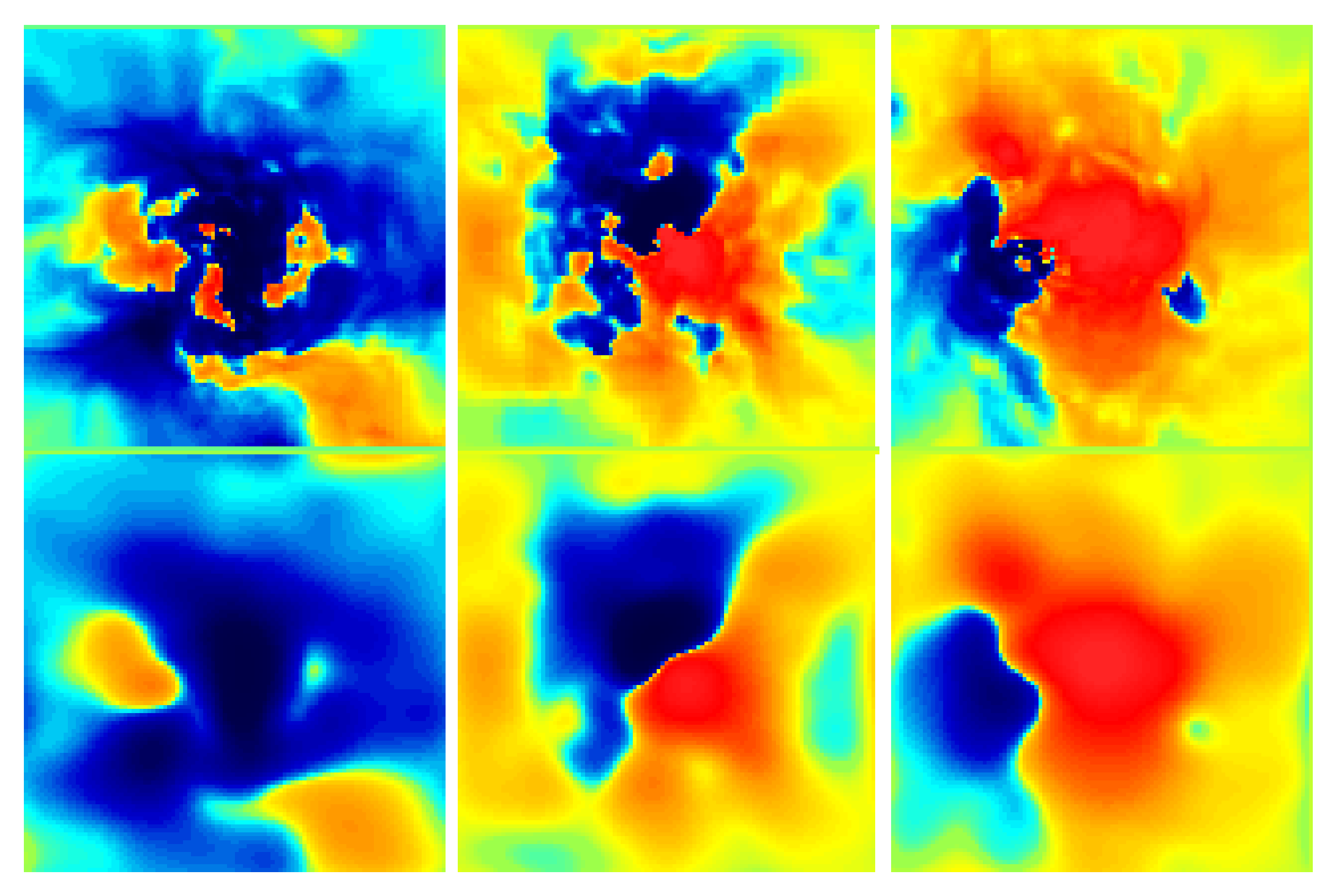} 
}
\vspace{0.15in}
\centerline{ 
\epsfysize=3.6truein  \epsffile{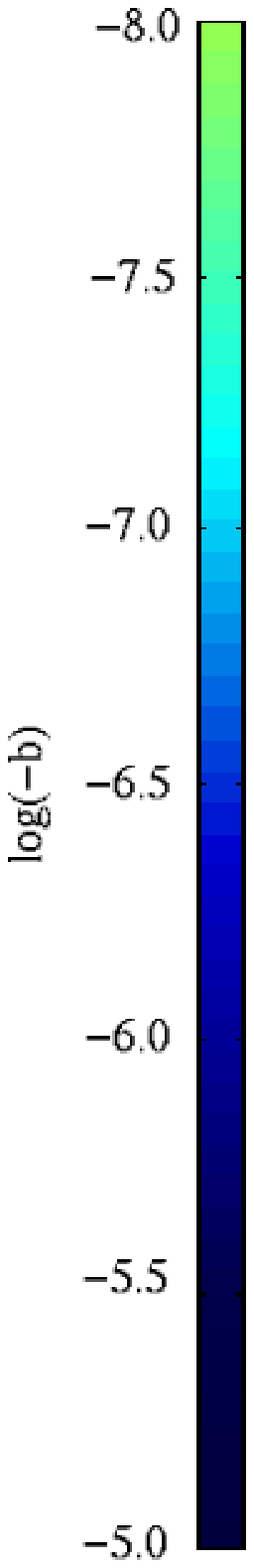} 
\epsfysize=3.6truein  \epsffile{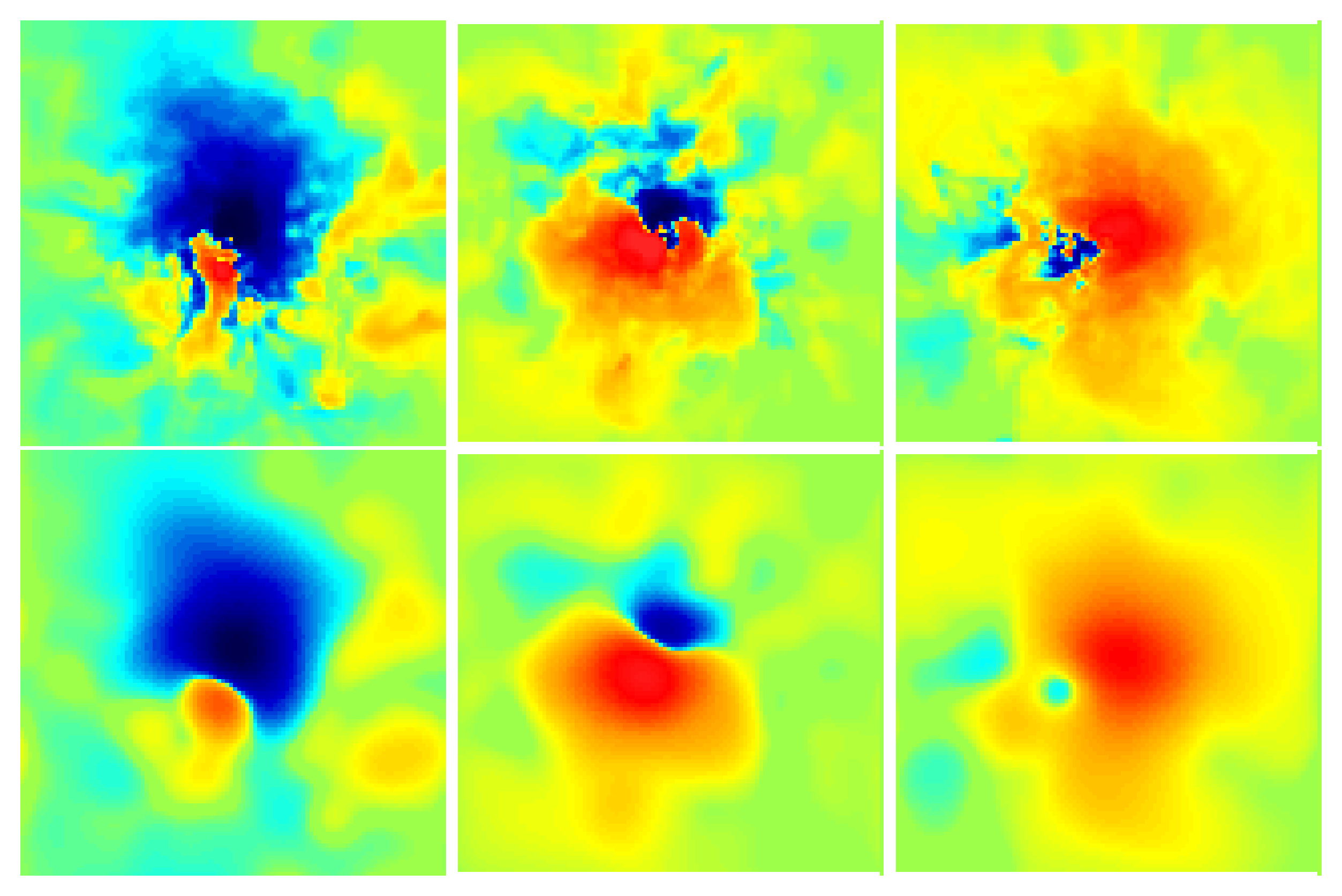} 
}
\vskip10pt
\figcaption{ \footnotesize
  {\it First and third rows:\/} noiseless kinematic SZ Effect (kSZ) maps of the
  simulated $\Lambda$CDM cluster at redshifts $z=0.43$ (1st row) and
  $z=0$ (2nd row) through three orthogonal projections.  {\it Second and fourth rows:\/} the kSZ maps shown in the top panels
  convolved with a 1' (FWHM) Gaussian beam, assuming the cluster is at
  a distance corresponding to redshift $z=0.1$.  The maps are
  color-coded on a $\log_{10}$ scale which is shown on the left hand
  side of the plots.  The comoving size of the region shown is 2$h^{-1}$ Mpc
  centered on the minimum of cluster potential, and the comoving pixel size is
  around 8$h^{-1}$ kpc.  Note that 1 arcminute
  corresponds to the physical scale 85 $h^{-1}$kpc at $z=0.1$.
\label{fig3}
\vskip2pt}
\end{figure*}

Figure~\ref{fig3} shows kSZ maps of the simulated cluster at two
different cosmic times (corresponding to redshifts $z=0.43$ and $z=0$)
in three orthogonal projections (the left panel corresponds to the
projection in Fig.~\ref{fig1}).  The top panels show the raw kSZ maps
of a simulated cluster, with the earlier time in the top row.  The
size of the region shown in the figure is 2$h^{-1}$ comoving Mpc and
the region is centered on the minimum of cluster potential.  The
comoving pixel size of the grid is $2\hMpc/256\approx 8\hkpc$, close
to the spatial resolution of the simulation.  The bottom panels show
the kSZ maps shown in the top panels convolved with a 1 arcminute FWHM
Gaussian beam, assuming a cluster angular diameter distance
corresponding to a redshift of $z=0.1$. One arcminute corresponds to
the physical scale 77.5$h^{-1}$ kpc at $z=0.1$ in the adopted
$\Lambda$CDM cosmology. The smoothed maps are heavily oversampled,
with a pixel size around a tenth of the beam width.  All maps are
color-coded on a $\log_{10}$ scale.

The high-resolution kSZ maps show a complex spatial structure of the
kSZ distortion signal associated with internal motion of cluster gas
and merging subclumps seen in Fig.~\ref{fig1}.  At $z=0.43$, the
cluster is in a dynamically active stage; two nearly equal-mass
sub-clusters have just undergone a slightly off-axis major merger and
created strong merger shocks.  In the top-left panel of
Fig.~\ref{fig3}, fast-moving gas ($\gtrsim 1000$ km/s) associated with
the merger shock fronts appears as a large positive kSZ signal at the
locations where the shock fronts are present, while the cluster as a
whole is moving away from the observer, resulting in an overall
negative kSZ signal when the raw kSZ signal is convolved with the
Gaussian beam.  In the top-middle panel of Fig.~\ref{fig3}, the merger
axis is aligned parallel to the line-of-sight direction, and the
merger shock fronts propagating in both directions of the merger axis
result in a dipole pattern in the kSZ signal.  At $z=0$, the cluster
is in a relatively relaxed state, but continues to accrete matter and
undergo minor mergers.  In the bottom-middle panel, the merging
subclump from Fig.~\ref{fig1} is moving parallel to the line of sight,
appearing as the strongly negative kSZ signal.  This behavior is
generic for galaxy clusters in our simulations: even epochs where the
cluster is ``relaxed'' minor merger(s) are typically present (see also
Fig.~\ref{fig2}). The ever-present internal motions within the cluster
produce prominent substructure in the kSZ maps.

\subsection{Estimating Radial Peculiar Velocity from kSZ Maps.}
\label{sec:est}

How do we estimate the cluster peculiar velocity from a kSZ map?  The
most straightforward approach is to define some aperture centered on
the cluster center (i.e., the peak of the thermal SZ signal or \xray
emission) and calculate the average velocity within the aperture.  As
we discuss below, this {\em aperture average velocity } is, in fact, a
largely unbiased estimator of a cluster's radial peculiar velocity,
and the error in each measurement will depend on the size and shape of
the aperture used to select the cluster region for the analysis.  For
accurate velocity measurements, the aperture must be chosen large
enough to encompass the bulk of the cluster region, and the optimal
aperture is matched to the virial radius of the cluster.

To construct a radial velocity estimator from the beam-smeared kSZ
maps, we assume that all of the gas in the cluster is moving with a
single radial bulk velocity.  To compute the aperture average velocity
from the maps, we first define a circular aperture centered on the
minimum of cluster potential and calculate the beam-smeared kSZ flux
$\langle b^B \rangle$ using all pixels within the aperture.  In
practice, the minimum of the gravitational potential is very close to
the maximum of the cluster X-ray emission or SZ signal.  Using the same
set of pixels, we also calculate the average beam-smeared electron
column density $\langle\tau^B\rangle$.  The estimated radial velocity
is then simply
\beq
v_r^{\rm est} = \frac{c}{\sigma_T} \: 
        \frac{\langle b^B \rangle}{\langle \tau^B \rangle}
\label{eq:v_est}
\eeq
Below, we consider three different apertures relative to the cluster
virial radius: $r_{vir}$/4, $r_{vir}$/2, and $r_{vir}$ and compute
$\langle\tau^B\rangle$ in two different ways.  In the context of a
simulation, it is simple to compute this quantity directly from the
line of sight average of optical depth.  Observationally, one needs to
estimate the thermal SZ effect and measure cluster gas temperature to
determine the beam-smeared column density, $\langle \tau^B \rangle = (
m_e c^2 /\sigma_T k_B) \times [\langle y^B \rangle / \langle T_e
\rangle_{n_e}^B ]$.  However, the density-weighted electron
temperature $\langle T_e \rangle_{n_e}^B$ is not an observable;
instead, the emission-weighted temperature of cluster gas $\langle T_X
\rangle$ from the \xray spectroscopy is often used.  If the
intracluster medium is isothermal throughout, then $\langle T_e
\rangle_{n_e}^B = \langle T_X \rangle$. Departures from isothermality
will give some systematic error in this approximation.  To mimic the
situation that will arise in the analysis of the actual observations,
we also compute $\langle\tau^B\rangle$ based on observables assuming
that the intracluster medium is isothermal throughout and evaluating
$\langle T_X \rangle$ within each aperture for the 0.5-2.0keV band,
using the \citet{raymond77} plasma emission model and assuming a
uniform metallicity of 0.3 solar.

\subsection{Definitions of Peculiar Velocity}
\label{sec:defpec}

To gauge the accuracy of estimated radial velocity from kSZ maps in
the previous section, we need to define a reference ``true'' radial
peculiar velocity. In the literature, the peculiar velocity is
typically defined to be the density-weighted average dark matter
velocity within a sphere of some characteristic cluster radius.  The
most natural outer boundary of a cluster is given by the virial
radius, but a fixed radius is also often used in the literature
\citep[e.g., Abell radius of 1.5 Mpc][]{bahcall94,croft94,colberg00}.
We define $v_r^{\rm vir}$ as the density-weighted average dark matter
velocity within the virial radius. In linear theory, the peculiar
velocity is defined to be the velocity of a density peak of dark
matter smoothed over 10$h^{-1}$Mpc.  In turn, the kSZ estimate of the
radial peculiar velocity is the density-weighted average gas velocity
well within the virial radius, as the kSZ signal is concentrated in
the central region of the cluster.

\myputfigure{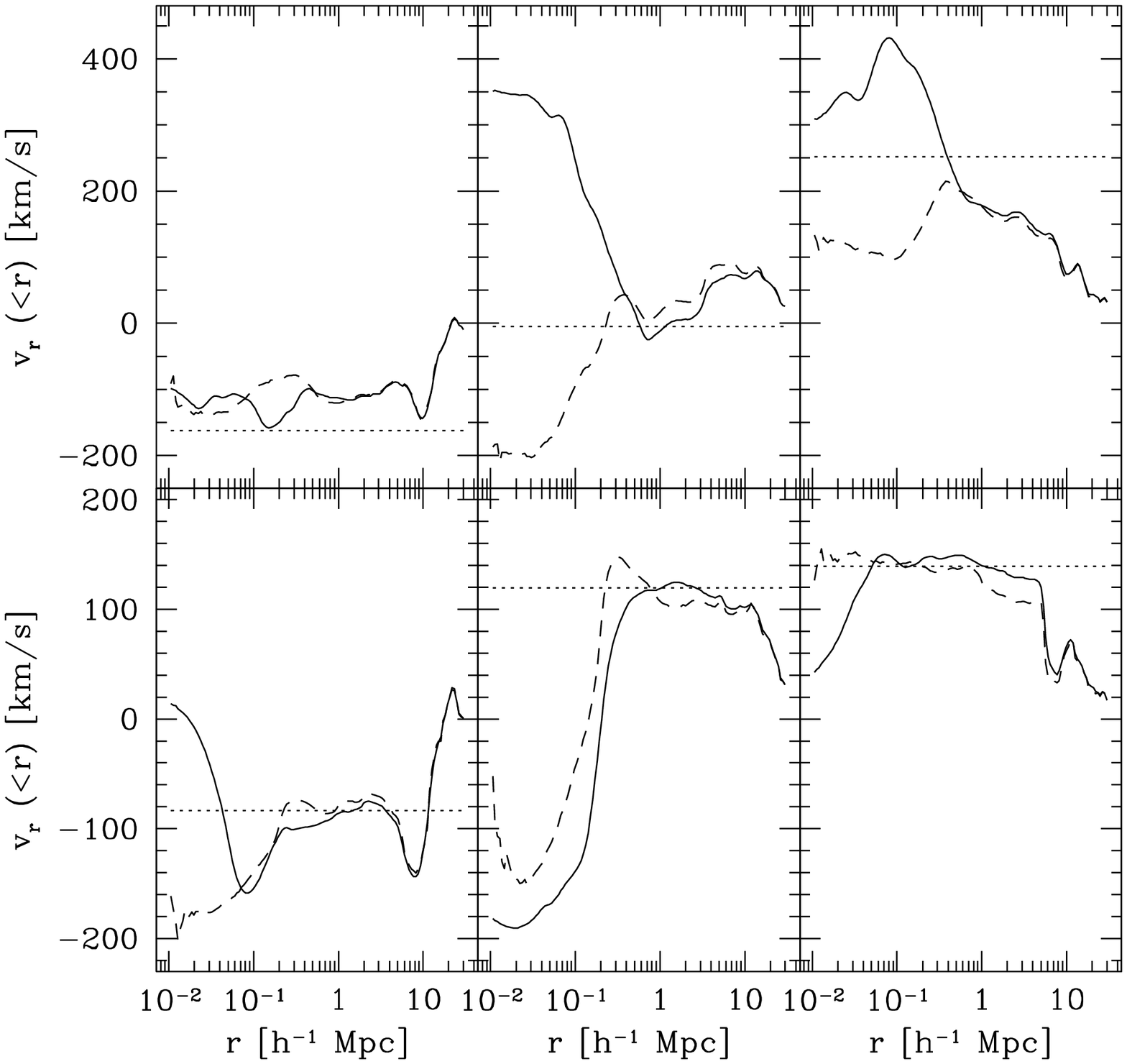}{3.2}{0.52}{-15}{-0}
\figcaption{
  \footnotesize The density-weighted average radial velocities of gas
  (solid) and dark matter (dashed) within a sphere centered on the
  minimum of cluster potential for clusters at $z=0.43$ (top) and
  $z=0$ (bottom) through the same three orthogonal projections (from
  left to right) shown in Fig.~\ref{fig3}.  Dotted lines show the
  estimated radial velocity of a cluster $v_r^{\rm est}$ from kSZ
  observations; $v_r^{\rm est}$ is the average velocity within the
  aperture matched to the virial radius of the cluster.  The
  density-weighted average velocities of gas and dark matter agree
  remarkably well beyond $\sim$500kpc (approximately half the cluster
  virial radius), particularly for the relaxed $z=0$ cluster; however,
  the discrepancies are significant in the central region of the
  cluster.
\label{fig4}
\vskip10pt
}

To illustrate how various definitions of the peculiar velocity are
related, we examine the density-weighted average gas and matter
velocity profiles as a function of radius.  Figure~\ref{fig4} shows
the density-weighted average radial velocities of gas (solid) and dark
matter (dashed) within a sphere centered on the minimum of the cluster
potential at redshifts $z=0.43$ (top) and $z=0$ (bottom)
through the same three orthogonal projections shown in
Fig.~\ref{fig3}. Dotted lines show the estimated radial velocity of a
cluster $v_r^{\rm est}$ from kSZ observations; here, $v_r^{\rm est}$
is the average velocity within the aperture matched to the virial
radius of the cluster.

The density-weighted average velocities of gas and dark matter agree
remarkably well beyond $\sim 500$ kpc (approximately half of the
virial radius of the cluster); however, the discrepancies are
significant in the central region of the cluster. The discrepancy
between the gas and dark matter velocity is most pronounced during the
merger when the internal gas velocities exceed 1000 km/s. This is not
surprising since dark matter and gas relax via different processes
(gas through shocks and dark matter via violent relaxation). Note also
that the peculiar velocity averaged within the virial radius can be
significantly different from the velocity averaged over a larger scale
(e.g., the difference with average velocity on 10$\hMpc$ scales
reaches $\gtrsim 50$\% in some cases).  The profiles show that,
contrary to common assumption, clusters do not move as solid bodies
and the choice of a peculiar velocity definition is somewhat
arbitrary.

\subsection{The Accuracy of kSZ Peculiar Velocity Estimates}
\label{sec:bias}

The top panel of Fig.~\ref{fig5} show a comparison of the estimated
radial velocities of a cluster from kSZ observations $v_r^{\rm est}$
and the density-weighted average velocity of dark matter within the
virial radius of the cluster $v_r^{\rm vir}$.  Here, we computed
$\langle\tau^B\rangle$ directly from the simulations.  We examine the
simulated cluster at nine different redshifts ranging between 0.66 and
0.00, with an interval of $\sim 0.8$ Gyr between redshifts, and
evaluate $v_r^{\rm est}$ and $v_r^{\rm vir}$ for three orthogonal
projections. The time interval is roughly equal to the dynamical time
of the cluster, so the various epochs provide largely independent
cluster configurations.  In the top panel, $v_r^{\rm est}$ is the
aperture average velocity within an aperture matched to the virial
radius $r_{\rm vir}$ (solid), $r_{\rm vir}$/2 (star) and $r_{vir}$/4
(circle) of the cluster.  The bottom-left panel of Fig.~\ref{fig5}
shows a histogram of $v_r^{\rm est}-v_r^{\rm vir}$ for an aperture
size matched to the virial radius $r_{vir}$ (solid) and $r_{vir}$/4
(dotted-hatched).

The figure shows that the {\em aperture average velocity } is indeed
an unbiased estimator of a radial peculiar velocity of the cluster.
When the aperture encompasses the entire cluster virial region, the
estimated velocity $v_r^{\rm est}$ agrees quite well with the true
peculiar velocity of a cluster $v_r^{\rm vir}$ (top panel) with a
small dispersion of $\lesssim 50$\kms (bottom panel).  When a smaller
aperture is used to select the inner region of the cluster for the
analysis, the estimated velocity is still unbiased, but the errors
become larger; the error is $\simeq 100$\kms if the aperture
encompasses a region inside the quarter of the virial radius.  The
error in each measurement will depend not only on the size but also
the shape of the aperture used to select the cluster region for the
analysis.  We consider a circular aperture for simplicity, but more
elaborate apertures can be used. However, any aperture that puts more
weight on the inner region of the cluster will likely lead to larger
errors due to the internal bulk motions of gas at a level of $\sim
200$\kms present in the central region of the cluster (see
Fig.~\ref{fig1} and Fig.~\ref{fig4}).

The kSZ velocity estimates are in good agreement with $v_r^{\rm vir}$
for the cluster, with virial radii of 1.14$\hMpc$ for $z=0.43$ and
1.26$\hMpc$ for $z=0$; the agreement is particularly good at the lower
redshift where the cluster is relaxed. This means that the various
large gas velocities average to the mean cluster velocity in general,
as expected from Fig.~\ref{fig2}.  However, our simulation suggests
that substructure which will produce a significant velocity error
occurs in a non-negligible fraction of clusters.  Comparison of
high-resolution kSZ maps and X-ray maps may identify such substructure
on a case-by-case basis; the extent to which the effect of subclumps
can be modelled from simulations remains to be seen. Without
modelling, systematic errors on the order of 50 to 100 km/s in kSZ
velocity estimates will occur in some fair number of clusters,
although statistically these errors should be unbiased since we are as
likely to observe subclumps falling in from either side of a cluster.
Maps of the kSZ effect with 1 arcminute resolution should clearly
identify the clusters with potentially serious errors from subclumps.

\myputfigure{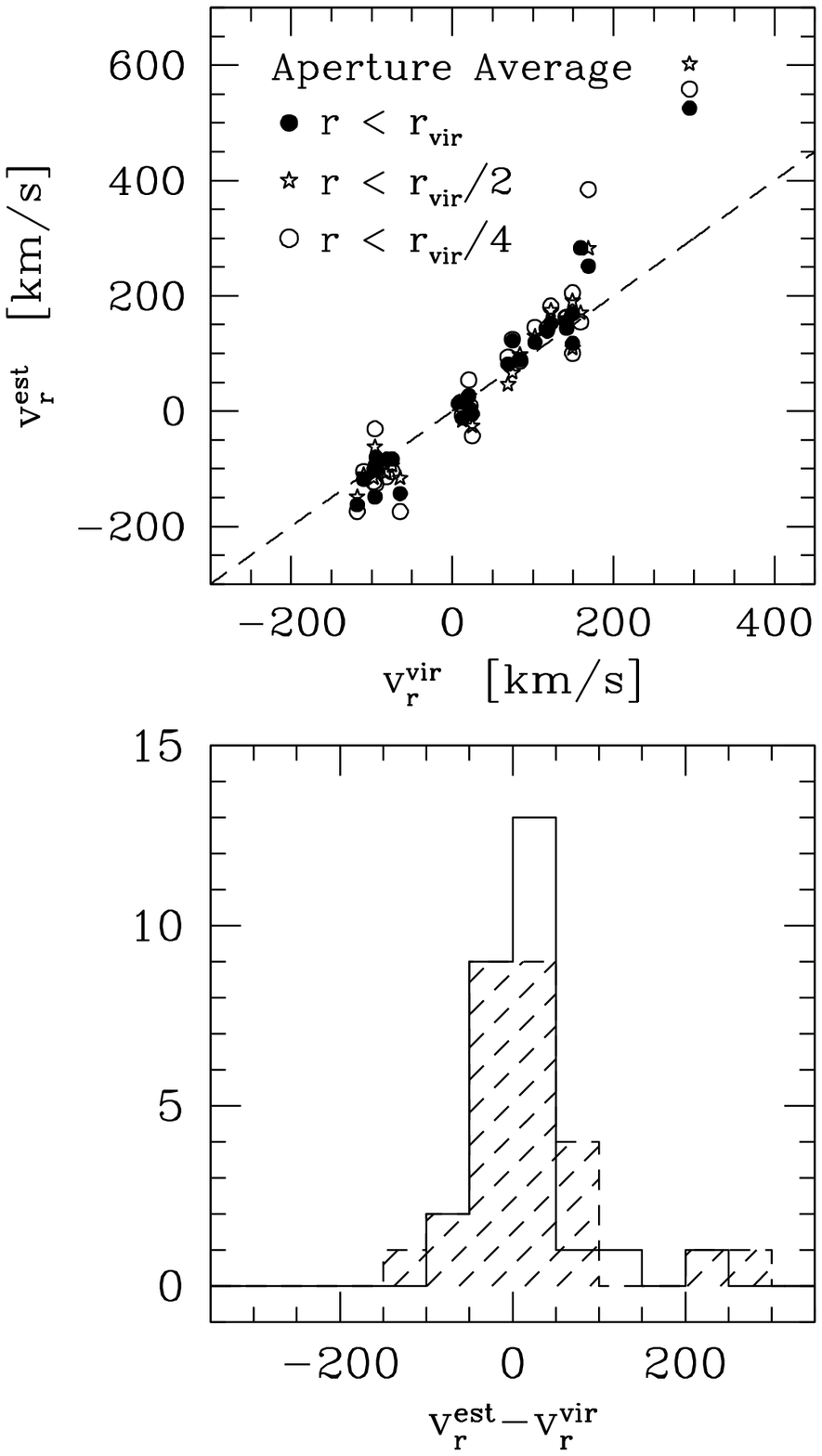}{3.2}{0.50}{-12}{+3} 
\figcaption{ \footnotesize {\it Top}: the estimated
  radial velocities of a cluster from kSZ observations, $v_r^{\rm
    est}$ (Eq.~(\ref{eq:v_est})), are compared with the
  density-weighted average velocity of dark matter within the virial
  radius of the cluster, $v_r^{\rm vir}$.  We use cluster outputs at
  nine different redshifts ranging between 0.66 and 0.00 separated by
  intervals of a cluster dynamical time ($\sim 0.8$ Gyr) and evaluate
  $v_r^{\rm est}$ and $v_r^{\rm vir}$ for three orthogonal
  projections.  In the top panel, $v_r^{\rm est}$ is the average
  velocity within an aperture matched to the virial radius $r_{\rm
    vir}$ (solid), $r_{\rm vir}$/2 (star) and $r_{\rm vir}$/4 (circle)
  of the cluster.  {\it Bottom}: a histogram of $v_r^{\rm
    est}-v_r^{\rm vir}$ for an aperture size matched to the virial
  radius $r_{vir}$ (left : solid) and $r_{vir}$/4 (left :
  dotted-hatched).  The aperture average radial velocity of a cluster
  from kSZ observations $v_r^{\rm est}$ is unbiased with a dispersion
  of $50-100$\kms, depending on the size of the aperture.
\label{fig5}
\vskip10pt
}

So far in the analysis, we used $\langle\tau^B\rangle$ derived
directly from the simulations.  Although this is a correct way to
compute this quantity in the context of the simulation, it is not a
direct observable.  In practice, one needs to derive this quantity
from thermal SZ effect and cluster gas temperature measurements.  To
mimic this situation, we performed the same analysis based on
observables as described in Sec.~\ref{sec:est}; the results show no
significant difference in scatter displayed in Fig.~\ref{fig5} and
Fig.~\ref{fig6}.  Our conclusions therefore are not very sensitive to
the details of the analysis or to the assumption of isothermality.

\myputfigure{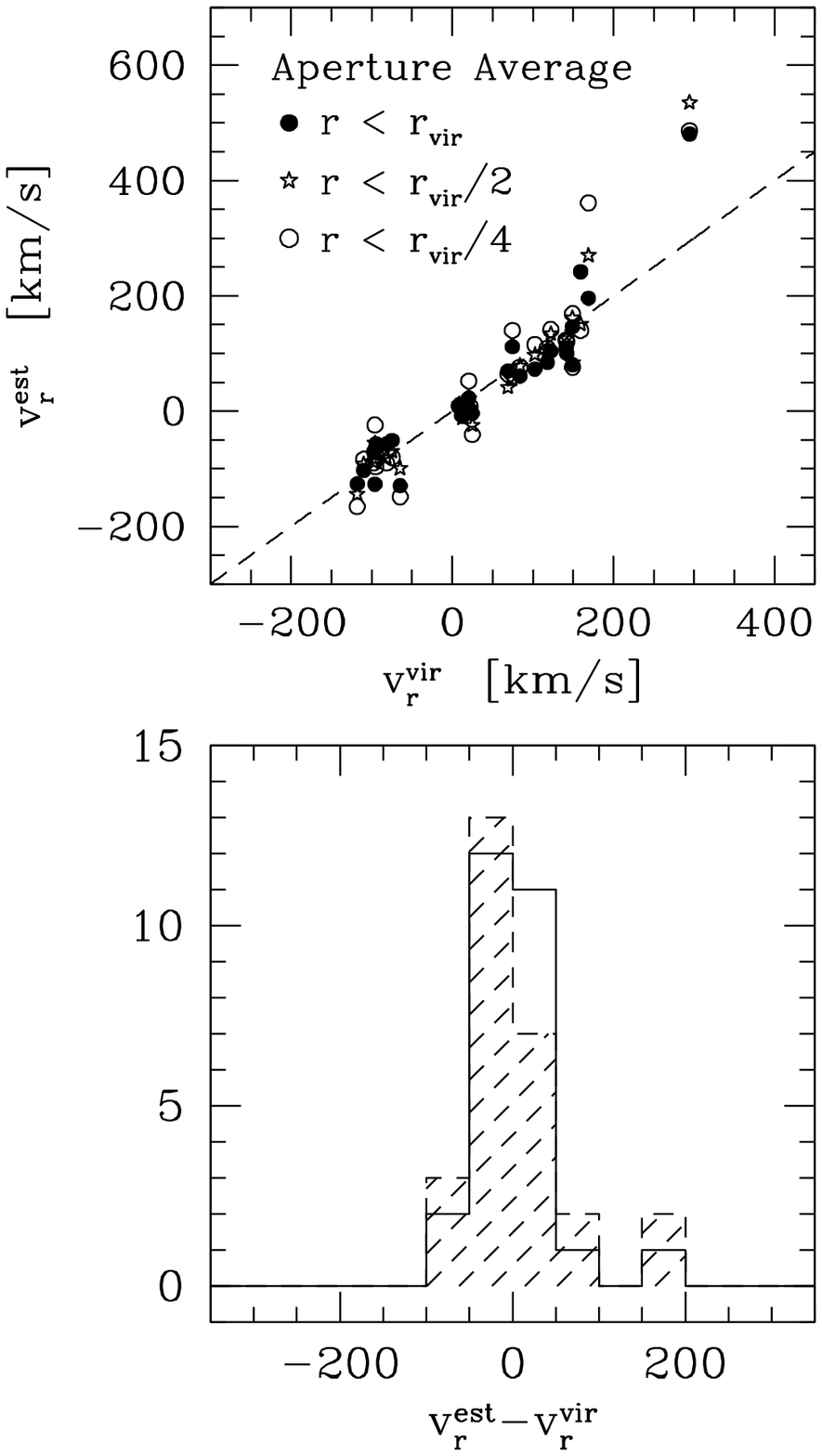}{3.2}{0.50}{-12}{+5} 
\figcaption{ \footnotesize 
Same as Fig.~\ref{fig5}, except that we are now computing
$\langle\tau^B\rangle$ using observable quantities as described in
Sec.~\ref{sec:est}, rather than deriving this quantity directly from
the simulations.  There is no significant difference between this
figure and Fig.~\ref{fig5}.
\label{fig6}
\vskip10pt
}

\myputfigure{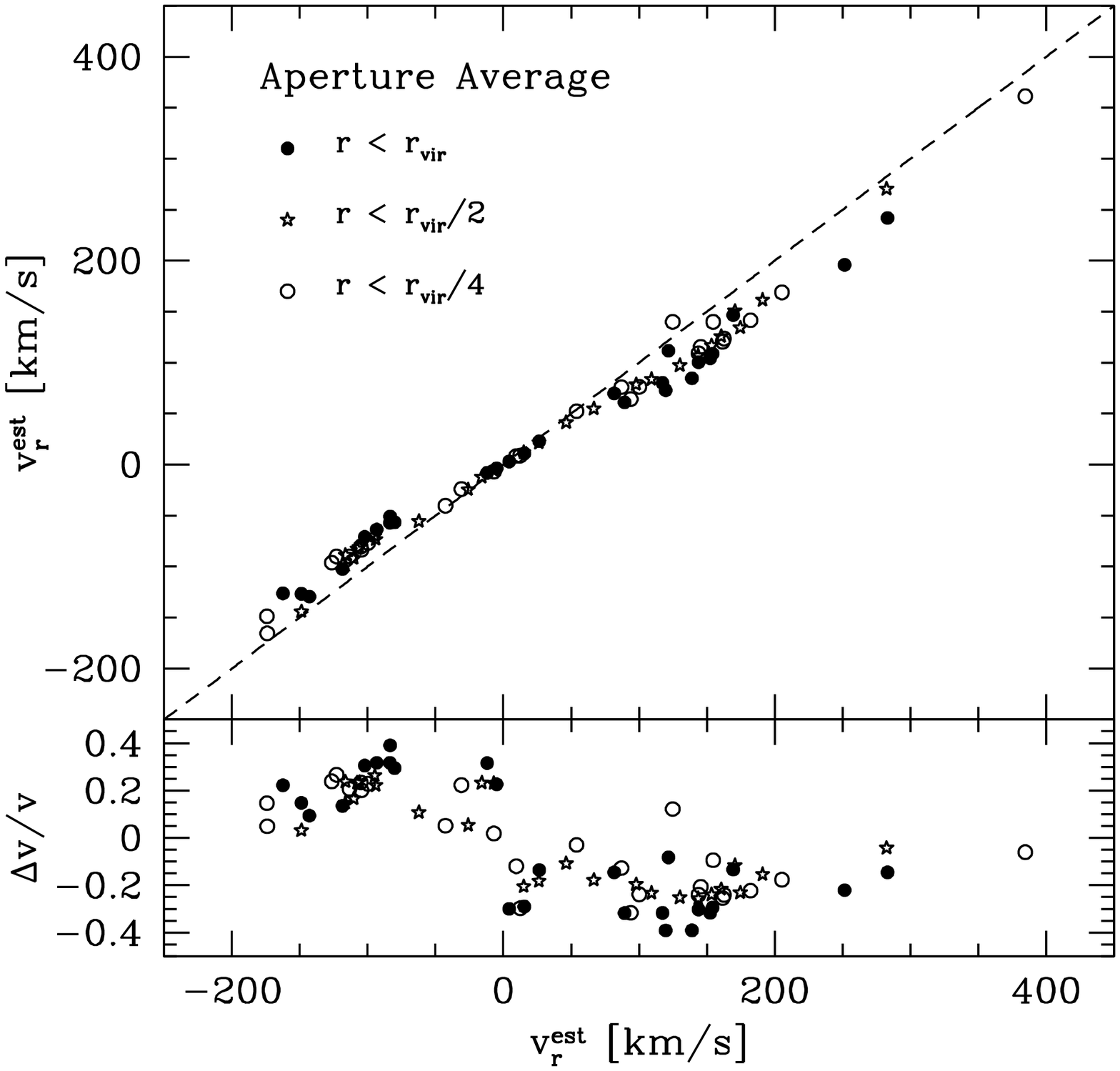}{3.2}{0.50}{-20}{-5} 
\figcaption{ \footnotesize 
{\it Top}: the radial peculiar velocities estimated using observables
($v^{est}_B$) are compared with the values using
$\langle\tau^B\rangle$ directly derived from the simulations
($v^{est}_A$) (see Sec.~\ref{sec:est}).  {\it Bottom}: the fractional
difference, $\Delta v/v = (v^{est}_B-v^{est}_A)/v^{est}_A$, is
plotted.  The magnitude of the radial peculiar velocity measurement is
systematically overestimated at a level of $\sim$20\% on average with
a scatter of $\sim$10\%, if we assume $\langle T_e \rangle_{n_e}^B =
\langle T_X \rangle$ within each aperture.  Here, the bias is caused
by a monotonically increasing temperature profile toward the cluster
center -- this is a generic prediction of an adiabatic cluster
simulation \citep[see e.g.,][]{frenk99}.
\label{fig7}
\vskip10pt
}

In the Fig.~\ref{fig7}, we compare the estimated radial peculiar
velocities derived using two different definitions of
$\langle\tau^B\rangle$.  The figure shows that the magnitude of the
radial peculiar velocity measurement is systematically overestimated
at a level of $\sim$20\% on average with a scatter of $\sim$10\%, if
we assume $\langle T_e \rangle_{n_e}^B = \langle T_X \rangle$ within
each aperture.  Here, the bias is caused by a monotonically increasing
temperature profile toward the cluster center -- this is a generic
prediction of adiabatic cluster simulations \citep[see
e.g.,][]{frenk99}.  The magnitude of this bias is comparable to the
magnitude of scatter due to internal motions of gas in cluster
cores, and the bias is systematic function of the estimated velocity
rather than random (see Sec.~\ref{sec:discuss} for a discussion about
a more realistic estimate of this bias.)

\section{Discussion and Conclusions}
\label{sec:discuss}

Using a high-resolution N-body+gasdynamic cluster simulation in a
$\Lambda$CDM cosmology, we have shown that

\newcounter{bean}

\begin{list}
{\roman{bean}}{ \usecounter{bean} \setlength{\parsep}{+0.03in}
\setlength{\leftmargin}{+0.15in} \setlength{\rightmargin}{+0.15in}}

\item{} 
Arcminute-scale kSZ measurements will provide a generally accurate
(with typical errors of $\lesssim 100$ km/s) tracer of cluster
peculiar velocities on scales of the cluster virial radius, but show
significant discrepancies on scales smaller than half of the virial
radius.

\item{}
In an ideal situation with an exact recovery of kSZ signal, the
velocity averaged over the kSZ map pixels inside a circular aperture
matched to the cluster virial region provides a statistically unbiased
estimate of a cluster's radial peculiar velocity with a small
dispersion of $\lesssim 50$\kms.  When analyzing an actual kSZ map
with noise, one may want to use a smaller aperture to include only the
region of the map with higher signal-to-noise; in such case, the
velocity measurement is still unbiased, but with larger errors.


\item{}
While accurate cluster velocity estimates might be obtained in a
statistical sense from kSZ estimates, any individual cluster can give
a significant velocity error due to random subclumps with velocities
along the line of sight. Comparisons of high-resolution kSZ and X-ray
maps can identify such subclumps, although it is unclear how
accurately their effect can be modelled on a case-by-case basis.

\end{list}

Our estimated error amplitude agrees with the conclusions of
\citet{holder02}.  Although our analysis shows that the accurate
measurements of radial peculiar velocity of clusters are possible,
careful observational and theoretical studies of potential biases are
required before we can use the kSZ peculiar velocity measurements to
extract cosmological information.

Observationally, one needs thermal SZ effect and cluster gas
temperature measurements to determine the beam-smeared column density,
$\langle \tau^B \rangle = ( m_e c^2 /\sigma_T k_B) \times [\langle y^B
\rangle / \langle T_e \rangle_{n_e}^B ]$.  However, the
density-weighted electron temperature $\langle T_e \rangle_{n_e}^B$ is
not an observable; instead, the emission-weighted temperature of
cluster gas $\langle T_X \rangle$ from the \xray spectroscopy is often
used.  If the intracluster medium is isothermal throughout, then
$\langle T_e \rangle_{n_e}^B = \langle T_X \rangle$.  However, this is
clearly an incorrect assumption: recent high-resolution \xray cluster
observations revealed the non-isothermality of the intracluster medium
\citep{markevitch98,peterson01,tamura01,degrandi02}.  Nevertheless, in
the absence of a theoretical model of temperature structure, the
isothermality assumption is likely an unavoidable one in the analysis
of the real observations, particularly for distant clusters where
detail temperature structure is difficult to study.  Clearly, such an
incorrect assumption could introduce an additional bias in the kSZ
measurements of the cluster peculiar velocity.  Using the
high-resolution adiabatic simulation used in this paper, we find that
the use of $\langle T_X \rangle$ leads to overestimate of the inferred
peculiar velocity measurement at the maximum level of $\sim$20\% due
to monotonically increasing temperature profile toward the cluster
center.  \citet{yoshikawa98}, on the other hand, reported that the use
of $\langle T_X \rangle$ leads to underestimate of similar magnitude
due to temperature drop in the center of the simulated cluster in
their rather low-resolution adiabatic cluster simulations.  However,
the real clusters \citep[e.g.,][]{degrandi02} and clusters simulated
with starformation and cooling \citep[e.g.,][]{valdarnini02} seem to
have more isothermal core than the simulated cluster analyzed here and
this bias will be correspondingly smaller.  Further analyses using
simulations with cooling and starformation are needed to evaluate the
effect, while reproducing the observed temperature profiles of
clusters.

The Sunyaev-Zeldovich effect is independent of redshift and unable to
distinguish between objects at different redshifts along the same line
of sight. While the chance of two massive galaxy clusters being
aligned is negligible \citep{voit01a}, the probability for random
alignment of a galaxy group ($M\simeq 10^{13}$ to $10^{14}$ $M_\sun$)
with a cluster will be somewhat higher and will appear like additional
substructure in the kSZ map. If the group has a large line-of-sight
velocity with the same sign as the cluster peculiar velocity, it will
be detectable in the same way as a high-velocity merging subclump, and
might be modelled out in the same way. Groups with smaller peculiar
velocities will bias the velocity determination and be harder to pick
out in the kSZ maps. This systematic error will be small, considering
that the high-velocity merging subclump in our simulation biases the
velocity determination by around 50 km/s and a random galaxy will give
a substantially smaller signal. This estimate is consistent with an
assessment of this effect by \citet{aghanim01} who argue that random
superposition of clusters with other objects along the line of sight
leads to typical {\it rms} velocity errors of $\lesssim 100$~km/s.  A
more accurate estimation of this bias requires modelling with
large-volume simulations.

We mention in passing that significant internal bulk velocities with
characteristic amplitudes of 100 to 300 km/sec will greatly complicate
measurements of cluster rotations with characteristic kSZ signals at
the $\mu$K level, as proposed by \citet{cooray02}. A realistic
assessment of this effect would need to use high-resolution simulated
clusters, such as the one studied here.

Measuring blackbody kSZ distortions at the few $\mu$K level of course
requires overcoming other systematics unrelated to the kSZ signal
itself. Most importantly, galaxy clusters will possess a dominant
thermal Sunyaev-Zeldovich distortion much larger than the kSZ signal.
This thermal signal can, to some extent, be extracted via its
departure from a blackbody spectrum using multiple frequency
measurements, but relativistic corrections \citep{rephaeli95,itoh01} or
departures from kinetic equilibrium \citep{blasi00} can complicate
this analysis \citep[see][]{holder02}.  Second, the blackbody kSZ
distortion must be separated from the blackbody primary temperature
fluctuations of the microwave background. This can be accomplished via
spatial filtering, since the primary microwave background fluctuations
possess little power on cluster scales. The errors in filtering will
give some unavoidable systematic error. Detailed simulations of the
filtering process were done by \citet{haehnelt96} and updated for
currently envisioned experimental capabilities by \citet{holder02}.

\citet{aghanim01} apply a simple filter to simulated Planck satellite
maps, finding velocity errors due to the filtering of 300 to 600 km/s
due to Planck's angular resolution.  Upcoming ground-based experiments
are likely to have higher sensitivity and resolution than Planck,
although with complications from atmospheric emission. The extent to
which the kSZ signal can be isolated from the background fluctuations
in arcminute resolution maps at high sensitivity needs to be studied
in more detail. We anticipate systematic errors comparable to those
from internal motions. Galactic dust emission and radio point sources
are unlikely to be major problems at 200 GHz frequencies and arcminute
angular scales, but lensed images of distant dusty galaxies could
contribute a significant confusion noise \citep{blain98} and may need
to be imaged at higher frequencies to extract the kSZ signal
accurately.

Our simulated galaxy cluster shows that, perhaps contrary to naive
expectation, the internal bulk flows in galaxy clusters do not present
an insurmountable source of systematic error for peculiar velocity
estimates based on the kinematic Sunyaev-Zeldovich effect.
Aperture-averaged velocity estimates are largely unbiased. 
High-velocity subclumps merging with the cluster along the line of
sight will induce systematic errors, but many of these clumps can be
identified as particularly bright kSZ signals, and can be removed via
comparison with X-ray maps. It is not unreasonable to hope that SZ
cluster surveys will eventually produce cluster peculiar velocity
catalogs with systematic velocity errors at a level of 50 km/s
independent of the cluster distance.  Velocity catalogs of such
accuracy would provide a powerful probe of the growth of structure in
the mildly nonlinear regime.


\acknowledgements 

We would like to thank Lloyd Knox for useful discussion and comments,
and Roman Juszkiewicz for asking questions prompting this study. The
work presented here was partially supported by NASA through a Hubble
Fellowship grant from the Space Telescope Science Institute, which is
operated by the Association of Universities for Research in Astronomy,
Inc., under NASA contract NAS5-26555 and by NSF through funding of the
Center for Cosmological Physics at the University of Chicago (NSF
PHY-0114422). DN thanks John Carlstrom for his support through NASA
LTSA grant NAG5--7986. AK is a Cottrell Scholar of the Research
Corporation and is supported by NASA's SARA program through 
grant NAG5-10110.



\end{document}

%% file: defs.tex
\newcommand{\beq}{\begin{equation}}
\newcommand{\eeq}{\end{equation}}

\def \xray {\hbox{X--ray} }

\def \logTd6 {\hbox{log$( T/6 \kev)$} }

\def\myputfigure#1#2#3#4#5%
{\vskip#5pt\makebox[0pt]{\hskip#2in
\includegraphics[width=#3\textwidth]{#1}}\vskip#4pt\hfill}

\def \hmpc       {{$h^{-1}$\rm\ Mpc}}

\def \kms       {\hbox{ km s$^{-1}$}}

\def \kev       {{\rm\ keV}}

\def \hmsol     {h^{-1}{\rm\ M}_\odot}
\def \hMpc      {h^{-1}{\rm\ Mpc}}
\def \hkpc      {h^{-1}{\rm\ kpc}}